\documentclass[12pt]{article}
\usepackage[usenames,dvipsnames,svgnames,table]{xcolor}
\usepackage{graphicx}
\usepackage[numbers]{natbib}
\usepackage{subfig}
\usepackage{epsfig}
\usepackage{rotating}
\usepackage{amsmath,amssymb}
\usepackage{url}
\usepackage{multirow}\textwidth6.5in \textheight9in \evensidemargin0pt \oddsidemargin0pt
\graphicspath{{./Figures/}}
\begin{document}

\begin{titlepage}
\begin{flushright}
	{CP3-14-02}
\end{flushright}
\vskip2cm
\begin{center}
{\Large Effective Field Theory Beyond the Standard Model} \\
\medskip
\bigskip\bigskip\bigskip\bigskip
\renewcommand{\thefootnote}{\fnsymbol{footnote}}
{\bf Scott Willenbrock}$^{1,2}$\footnote{e-mail: willen@illinois.edu}
 and {\bf Cen Zhang}$^3$\footnote{e-mail: cen.zhang@uclouvain.be} \\
 \setcounter{footnote}{0}
\bigskip\bigskip\bigskip
$^1$Department of Physics, University of Illinois at Urbana-Champaign \\
1110 West Green Street, Urbana, IL  61801 \\
\bigskip
$^2$Fermi National Accelerator Laboratory, P.\ O.\ Box 500, Batavia, IL  60510 \\
\bigskip
$^3$Centre for Cosmology, Particle Physics and Phenomenology (CP3) \\
Universit\'e Catholique de Louvain, B-1348 Louvain-la-Neuve, Belgium
\end{center}

\bigskip\bigskip\bigskip

\begin{abstract}

We review the effective field theory approach to physics beyond the Standard
Model using dimension-six operators. Topics include the choice of operator
basis, electroweak boson pair production, precision electroweak physics
(including one-loop contributions), and Higgs physics. By measuring the
coefficients of dimension-six operators with good accuracy, we can hope to infer
some or all of the features of the theory that lies beyond the Standard Model.

\end{abstract}
%
%
%
\end{titlepage}

\newpage
\tableofcontents
\newpage

\section{Introduction}
\label{sec:intro}

There are two methods to search for physics beyond the Standard Model of
particle physics.  One is to search for new particles.  The other is to search
for new interactions of known particles.  The latter is the subject of this
article.

There is a long history of physicists supplementing established theories with
new interactions, followed by experiments to measure or bound these
interactions.  In this article we concentrate on particle physics at the energy
frontier, in which the established theory is the Standard Model, including the
Higgs particle that was discovered in July 2012 by the ATLAS and CMS experiments
at the CERN Large Hadron Collider from proton-proton collisions at 7 and 8 TeV
\cite{Aad:2012tfa,Chatrchyan:2012ufa}.  This particle has been studied in some
detail and it behaves as the Standard Model Higgs particle to a good
approximation, so we will treat it thusly. At the time of this writing there
have been no particles found beyond the Standard Model, as we await
proton-proton collisions at 13 - 14 TeV.

The first question that arises is how one should go about parameterizing the
hypothesized new interactions.  However one extends the Standard Model, it
should not do violence with the framework of the Standard Model itself.  Since
the Standard Model is a quantum field theory, the extension should also be a
quantum field theory and respect all the axioms of quantum field theory.

The new interactions should respect the $SU(3)_C\times SU(2)_L\times U(1)_Y$
gauge symmetries of the Standard Model.  While the electroweak interaction is
spontaneously broken at energies below the mass of the Higgs boson, $m_h \approx
126$ GeV, it is unbroken above this energy.  Thus the gauge symmetries of the
Standard Model are unbroken at the energy frontier.

We will assume that the new interactions decouple from the Standard Model in the
limit that the energy scale that characterizes the new interactions goes to
infinity. This is not true of all theories beyond the Standard Model, but it is
a good working assumption, and is consistent with what we know at present.

Many sectors of the Standard Model have been tested with an accuracy that goes
beyond the leading order in perturbation theory.  We therefore desire that any
new interactions that are introduced allow for unambiguous calculations beyond
leading order.  Ideally one should be able to calculate any process at any order
in both the Standard Model interactions and the new interactions.

All of the above properties are satisfied by an effective field theory
\cite{Weinberg:1978kz,Weinberg:1980wa,Georgi:1994qn}.  As the name implies, it
is a quantum field theory, and it is straightforward to implement the gauge
symmetries of the Standard Model.  It carries with it an energy scale,
$\Lambda$, that we interpret as the scale of new physics and that we assume to
be greater than the Higgs mass.  New interactions are constructed from Standard
Model fields and have a coefficient proportional to an inverse power of
$\Lambda$; thus the Standard Model is recovered in the limit $\Lambda \to
\infty$.  The new interactions are compatible with the calculation of Standard
Model radiative corrections.  The new interactions may be calculated to any
desired order in $1/\Lambda$, with the caveat that the introduction of
additional interactions may be necessary at each order in $1/\Lambda$.

A benefit of thinking of the Standard Model as an effective field theory is that
it explains some of its mysteries.  Why does the Standard Model contain only
renormalizable interactions, and not the more complicated nonrenormalizable
interactions?  The answer is that the nonrenormalizable interactions have
coefficients proportional to inverse powers of the scale of new physics,
$\Lambda$, and are therefore suppressed at energies less than $\Lambda$.  If one
writes down every interaction constructed from Standard Model fields with
coefficients that are dimensionless or of positive mass dimension, one arrives
at the Standard Model.

Another benefit of thinking of the Standard Model as an effective field theory
is that it gives some guidance as to what new interactions to look for.  One
would expect the dominant new interactions to be the ones least suppressed by
inverse powers of $\Lambda$.  In the absence of any restrictions, the new
interactions that are the least suppressed are proportional to the first power
of $1/\Lambda$.  There is only one such interaction (for one generation of
fermions), and it gives rise to Majorana masses for neutrinos
\cite{Weinberg:1979sa}.  Since our focus is on the energy frontier, we will not
discuss this interaction any further.

The interactions that are suppressed by $1/\Lambda^2$ come in two classes.
First are the interactions that violate baryon number ($B$) and lepton number
($L$).  There are 5 such interactions (for one generation)
\cite{Weinberg:1979sa,Wilczek:1979hc,Abbott:1980zj}, and the ones involving the
light quarks and leptons are highly constrained by experiments such as proton
decay.  Second are the interactions that respect $B$ and $L$, which are less
constrained.  There are a staggering number of these interactions: the number of
independent interactions of this type is 59 (for one generation)
\cite{Grzadkowski:2010es}.  By the same counting, the number of interactions in
the Standard Model is only 14 (for one generation).  Thus we gain a renewed
appreciation for the simplicity of the Standard Model.

The interaction that gives rise to Majorana masses for neutrinos violates $L$ by
two units.  In fact, all interactions that are proportional to an odd power of
$1/\Lambda$ violate $B$ and/or $L$ \cite{Degrande:2012wf}.  We will restrict our
attention to the interactions that conserve $B$ and $L$, since these are the
least constrained and therefore the most likely to be relevant to experiments at
the energy frontier.

The effective field theory of the Standard Model that conserves $B$ and $L$ is thus
\begin{equation}
{\cal L} = {\cal L}_{SM} + \sum_i \frac{c_i}{\Lambda^2}{\cal O}_i + \cdots
\;,
\label{eq:L}
\end{equation}
where $c_i$ is a dimensionless coefficient\footnote
{These are often called Wilson coefficients to honor Ken Wilson's
pioneering work on effective field theory.}
and ${\cal O}_i$ is an operator\footnote{If $O_i$ is not hermitian, its
hermitian conjugate should be added to Eq.~(\ref{eq:L}).} constructed from
Standard Model fields.  Since a Lagrangian is of mass dimension four, the
operators ${\cal O}_i$ are of mass dimension six.  The ellipsis refers to
operators of dimension eight, ten, {etc}.  The expectation is that the
leading effects of new physics will be represented by the dimension-six
operators, since they are the least suppressed.

An example of an effective field theory is given by a new $Z^\prime$ boson that
couples to Standard Model fermions.  At energies below the mass of the
$Z^\prime$, one will not see the new particle directly, but rather a new
four-fermion interaction of Standard Model fermions.  Since fermion fields are
of mass dimension 3/2, this is represented by a dimension-six operator.  The
$Z^\prime$ propagator contributes a factor of $1/M_{Z^\prime}^2$ to the
interaction, so the scale of new physics, $\Lambda$, is the mass of the
$Z^\prime$.  The dimensionless coefficient of the dimension-six operator, $c_i$,
is the product of the $Z^\prime$ couplings to the Standard Model fermions.  The
reader may recognize that the above example is analogous to the Fermi theory of
the weak interaction, which is the effective field theory of the weak
interaction at energies below the $W$ mass.

This example illustrates that an effective field theory is not intended to be
valid to arbitrarily high energies, but only up to the scale of new physics,
$\Lambda$.  Beyond that energy, the new particles must be included explicitly.
One may then construct a new effective field theory from the Standard Model
particles and the new particles that have been discovered.

At this time there is no established evidence for the presence of any
dimension-six operators.  One may instead use the world's data to place bounds
on the coefficients of these operators.  In this way we quantify the accuracy
with which the new interactions are excluded.  The advantage of this approach is
that it does not make reference to any particular experiment.  Different
experiments place bounds on different combinations of dimension-six operators.
If every analysis uses the same approach, we are able to combine the world's
data in a consistent way.

If we were fortunate enough to find that the coefficients of some of the
dimension-six operators were nonzero, it would amount to the discovery of new
physics.  By measuring the coefficients with good accuracy, one could hope to
infer some or all of the features of the theory that underlies the effective
field theory.  There is an historical precedent in the theory of the weak
interaction.  Fermi's original theory of the weak interaction,  developed
shortly after the discovery of the neutron in 1932, was based on vector
currents; it took many years of experimentation to establish that the true
theory involved vector minus axial-vector ($V-A$) currents
\cite{Sudarshan:1958vf,Feynman:1958ty}.  This ultimately led to the introduction
of the $SU(2)_L$ gauge group as a fundamental aspect of the electroweak theory
\cite{Glashow:1961tr,Weinberg:1967tq,Salam:1968rm}.

The coefficients of dimension-six operators are dimensionful, $c_i/\Lambda^2$.
Unfortunately, measuring these coefficients does not reveal the scale of new
physics, $\Lambda$, only the ratio $c_i/\Lambda^2$. For example, the $V-A$
effective theory of the weak interaction by itself did not reveal the mass of
the $W$ boson, only the ratio $G_F/\sqrt 2 = g^2/8m_W^2$.  Searches were
performed for $W$ bosons as light as a few GeV \cite{Bernardini:1964im}.
However, once the $SU(2)_L\times U(1)_Y$ theory was proposed and weak neutral
currents were measured, there was enough information to infer the mass of the
$W$ (and $Z$) boson.  This is because it was possible to derive the weak
coupling $g$ from the electromagnetic coupling $e$ via $e = g \sin\theta_W$,
with $\sin\theta_W$ extracted from weak neutral current experiments.  Thus once
one has an underlying theory in mind, it may be possible to infer the scale of
new physics, $\Lambda$.

\section{Operator basis}
\label{sec:basis}

If one writes down all dimension-six operators that are constructed from
Standard Model fields and that respect the Standard Model gauge symmetries (as
well as $B$ and $L$), one arrives at a list that is longer than the 59 operators
mentioned previously.  However, one finds that many of these operators are
redundant; they are equivalent to a linear combination of other operators.
These linear relations correspond to the Standard Model equations of motion or
other identities \cite{Politzer:1980me}. This redundancy is a feature of
dimension-six operators that is unfamiliar from the Standard Model, where one
does not encounter redundant operators.

Because of this redundancy, there is a great deal of flexibility in which set of
59 operators to use.  Any set of 59 independent operators constitutes a good
basis.  However, there is no physically preferred basis: any basis can be used
to describe the data.  A particular experimental measurement generally depends
on only a few of the 59 operators in a given basis.

One can place a bound on the coefficient of a particular operator by assuming
that all the other operators in that basis have vanishing coefficients.
However, this is an {ad hoc} assumption, unless one has in mind an
underlying theory that produces this particular pattern of coefficients.  In the
absence of an underlying theory, one should include every dimension-six operator
that contributes to the calculation of a physical process.  Thus each
experimental measurement will generally bound a set of dimension-six operators.

There are several operator bases that are currently popular, and we characterize
them as follows:
\begin{itemize}
\item The first attempt to construct a complete basis goes back to
  Ref.~\cite{Buchmuller:1985jz}, which contains 80 dimension-six operators (for
  one generation). Gradually it was discovered that some of these operators are
  redundant \cite{Grzadkowski:2003tf,Fox:2007in,AguilarSaavedra:2008zc},
  eventually leading to the basis of 59 operators in
  Ref.~\cite{Grzadkowski:2010es}.  Thus the basis of
  Ref.~\cite{Grzadkowski:2010es} may be viewed as the descendant of the first
  attempt at constructing a complete basis in Ref.~\cite{Buchmuller:1985jz}, and
  is thus of historical significance, at the very least.  We will refer to this
  as the BW basis, with apologies to those who contributed to its development.
\item A basis that maximizes the number of operators involving only Higgs and
  electroweak gauge bosons was constructed in Ref.~\cite{Hagiwara:1993ck}.
  Since this paper did not discuss operators involving fermions, it was not
  intended to be a complete basis.  Nevertheless, it is widely used for studies
  of Higgs and weak boson physics, and can be extended to a complete basis by
  adding operators involving fermions. We will refer to this as the HISZ basis.
  When extended to include operators involving fermions, one may choose to
  eliminate some of the HISZ operators via equations of motion.
\item A third class of bases began with Ref.~\cite{Giudice:2007fh}, and has been
  developed into a complete basis
  \cite{Masso:2012eq,Contino:2013kra,Elias-Miro:2013gya,Elias-Miro:2013mua,Pomarol:2013zra},
  motivated in part by the discovery of the Higgs boson.  We will refer to this
  class of bases as the GGPR basis, again with apologies to those who
  contributed to its development.\footnote{This basis is sometimes referred to
    by the acronym SILH (Stongly Interacting Light Higgs), which is the title of
    the GGPR paper \cite{Giudice:2007fh}.  We prefer to avoid this label as this
  basis is of general applicability.}
\end{itemize}
Other bases are also in use, but we do not attempt to summarize them. For
example, the basis of Ref.~\cite{Elias-Miro:2013eta} is a cross between the BW
basis and the GGPR basis.

We show in Table~\ref{tab:bases} the $CP$-even operators containing only
electroweak boson fields in these three bases.  If the operators look strange
and unfamiliar, it is because they are: after all, they are not present in the
Standard Model, so we are not accustomed to them.  It takes some time to get
used to them, and it does not help that there are three different bases
presently in use, with three different notations and normalizations.  For
example, $O_W$ represents three completely different operators in the three
bases.  We will meet some of these operators as we proceed.

\renewcommand{\O}{\mathcal{O}} 
\newcommand{\tr}{\mathrm{Tr}} 
\newcommand{\ha}{\varphi} 
\newcommand{\hb}{\Phi} 
\newcommand{\hc}{H} 
\begin{table}
\caption{The $CP$-even operators containing only electroweak boson fields.$^a$
  \label{tab:bases}}
  \begin{equation}
    \begin{array}{|l|l|l|}
      \hline
      \multicolumn{1}{|c|}{\mathrm{BW}^b} &
      \multicolumn{1}{|c|}{\mathrm{HISZ}^c} &
      \multicolumn{1}{|c|}{\mathrm{GGPR}^d}
      \\\hline
      \O_{W}=\epsilon^{IJK}W_{\mu}^{I\nu}W_{\nu}^{J\rho}W_{\rho}^{K\mu} &
      \O_{WWW}=\tr[\hat{W}_{\mu\nu}\hat{W}^{\nu\rho}\hat{W}_\rho^\mu] &
      \O_{3W}=\frac{1}{3!}g\epsilon_{abc}W_{\mu}^{a\nu}W_{\nu\rho}^bW^{c\rho\mu}
      \\\hline
      \O_{\varphi W}=\ha^\dagger\ha W_{\mu\nu}^I W^{I\mu\nu} &
      \O_{WW}=\hb^\dagger \hat{W}_{\mu\nu} \hat{W}^{\mu\nu} \hb &
\textrm{---}
      \\\hline
      \O_{\varphi B}=\ha^\dagger\ha B_{\mu\nu} B^{\mu\nu} &
      \O_{BB}=\hb^\dagger \hat{B}_{\mu\nu} \hat{B}^{\mu\nu} \hb &
      \O_{BB}=g'^2|\hc|^2B_{\mu\nu}B^{\mu\nu}
      \\\hline
      \O_{\varphi WB}=\ha^\dagger\sigma^I\ha W_{\mu\nu}^I B^{\mu\nu} &
      \O_{BW}=\hb^\dagger\hat{B}_{\mu\nu}\hat{W}^{\mu\nu}\hb &
\textrm{---}
      \\\hline
\textrm{---}       &
      \O_W=(D_\mu\hb)^\dagger \hat{W}^{\mu\nu} (D_\nu\hb) &
      \O_{HW}=ig(D^\mu\hc)^\dagger\sigma^a(D^\nu\hc)W^a_{\mu\nu}
      \\\hline
\textrm{---}       &
      \O_B=(D_\mu\hb)^\dagger \hat{B}^{\mu\nu} (D_\nu\hb) &
      \O_{HB}=ig'(D^\mu\hc)^\dagger(D^\nu\hc)B_{\mu\nu}
      \\\hline
\textrm{---}       &
      \O_{DW}=\tr\left( [D_\mu,\hat{W}_{\nu\rho}][D^\mu,\hat{W}^{\nu\rho}] \right) &
      \O_{2W}=-\frac{1}{2}\left( D^\mu W^a_{\mu\nu} \right)^2
      \\\hline
\textrm{---}       &
      \O_{DB}=-\frac{g'^2}{2}(\partial_\mu B_{\nu\rho})(\partial^\mu{B}^{\nu\rho}) &
      \O_{2B}=-\frac{1}{2}\left( \partial^\mu B_{\mu\nu} \right)^2
      \\\hline
 \textrm{---}      &
 \textrm{---}      &
      \O_{W}=\frac{ig}{2}\left( \hc^\dagger\sigma^a \overleftrightarrow{D}^\mu \hc\right)
      D^\nu W^a_{\mu\nu}
      \\\hline
\textrm{---}       &
\textrm{---}       &
      \O_{B}=\frac{ig'}{2}\left( \hc^\dagger \overleftrightarrow{D}^\mu \hc\right)
      \partial^\nu B_{\mu\nu}
      \\\hline
      \O_{\varphi D}=\left( \ha^\dagger D^\mu\ha \right)^*
      \left( \ha^\dagger D_\mu\ha \right) &
      \O_{\Phi,1}=(D_\mu\hb)^\dagger\hb \hb^\dagger(D^\mu\hb)  &
      \O_{T}=\frac{1}{2}\left(\hc^\dagger\overleftrightarrow{D}_\mu\hc\right)^2
      \\\hline
      \O_{\varphi\Box}=(\ha^\dagger\ha)\Box(\ha^\dagger\ha) &
      \O_{\Phi,2}=\frac{1}{2}\partial_\mu(\hb^\dagger\hb)\partial^\mu(\hb^\dagger\hb)  &
      \O_{H}=\frac{1}{2}(\partial^\mu|\hc|^2)^2
      \\\hline
      \O_{\varphi}=(\ha^\dagger\ha)^3 &
      \O_{\Phi,3}=\frac{1}{3}\left( \hb^\dagger\hb \right)^3  &
      \O_6=\lambda|\hc|^6
      \\\hline
\textrm{---}       &
      \O_{\Phi,4}=(D_\mu\hb)^\dagger(D^\mu\hb)(\hb^\dagger\hb)  &
\textrm{---}
      \\\hline
    \end{array}
    \label{}
    \nonumber
  \end{equation}
  {\footnotesize
  $^a$ Operators of three popular bases of dimension-six operators are listed.
  The operators in each row are either identical (up to normalization) or similar.\\
  $^b$The notation and normalization of the BW basis \cite{Buchmuller:1985jz} are from
  Ref.~\cite{Grzadkowski:2010es}.\\
  $^c$The notation and normalization of the HISZ basis \cite{Hagiwara:1993ck} are from
  Refs.~\cite{Hagiwara:1993ck,Hagiwara:1996kf}.\\
  $^d$The notation and normalization of the GGPR basis \cite{Giudice:2007fh} are from Ref.~\cite{Elias-Miro:2013mua}.}
\end{table}

There are a large number of dimension-six operators containing fermions, and we
do not list them here.  They tend to be nearly the same in the three popular
bases, although there are differences that can be important.  For example, the
basis used in Ref.~\cite{Pomarol:2013zra}, which is in the GGPR class of bases,
uses equations of motion to eliminate two of the operators containing fermions
present in the BW basis.

Some operators are $CP$-odd.  At tree level, they interfere with the Standard
Model only if the observable is  constructed using triple product correlation
of momenta and/or spins, and thus the effect can only be observed in processes
where there are at least four independent momenta and/or spins that can be
measured.  We do not consider $CP$-odd operators in this review.

If the underlying theory is a weakly-coupled renormalizable gauge theory, it is
possible to classify dimension-six operators as being potentially generated at
tree level or at one loop \cite{Arzt:1994gp}.  This classification is cleanest in
a basis containing the maximum number of potentially-tree-generated operators
\cite{Einhorn:2013kja}.  Both the BW basis and the GGPR basis satisfy this
criteria, while the HISZ basis does not.  It has been argued that this
classification also applies to strongly-coupled underlying theories that are
minimally coupled \cite{Grojean:2006nn}, although the principle of minimal
coupling (which is not a principle of the Standard Model) has been criticized
\cite{Jenkins:2013fya}. 

Our own view is that one should fit the data with dimension-six operators
without regard to their classification.  If the data indicate that some
dimension-six operators have non-vanishing coefficients, the classification of
the operators may help in unraveling the underlying theory.

\section{Weak boson pair production}
\label{sec:WW}

As an example of the application of effective field theory to a physical
process, let's consider weak boson pair production, specifically $W^+W^-$
production.  This process has been measured at LEP II via $e^+e^-\to W^+W^-$ and
at the Tevatron and LHC via $q\bar q\to W^+W^-$.

Dimension-six operators affect this process in two different ways.  First, they
can affect the processes from which the input parameters $\alpha$, $G_F$, and
$M_Z$ are derived. Second, dimension-six operators can contribute directly to
this process.  A given operator may even contribute in both ways.

\begin{figure}[tb]
  \begin{center}
    \includegraphics{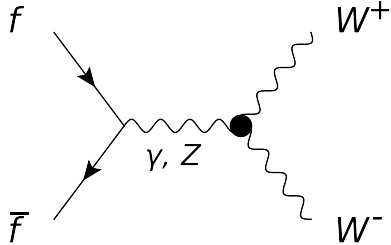}
  \end{center}
  \caption{Feynman diagram for $f\bar{f}\rightarrow W^+W^-$.  The dot
  represents an interaction arising from a dimension-six operator.}
  \label{fig:1}
\end{figure}

Let's consider the dimension-six operator
\begin{equation}
{\cal O}_{WWW}= \tr[\hat{W}_{\mu\nu}\hat{W}^{\nu\rho}\hat{W}_\rho^\mu]
\;,
\label{eq:WWW}\end{equation}
where
\begin{equation}
\hat{W}_{\mu\nu}= \frac{i}{2}g\sigma^a(\partial_\mu W_\nu^a-\partial_\nu
W_\mu^a-g\epsilon^{abc}W_\mu^b W_\nu^c)
\end{equation}
is the $SU(2)_L$ field strength tensor.  This operator is present in all three
bases of Table~\ref{tab:bases}, although with different notation and
normalization; we have adopted the HISZ conventions for this discussion.  This
operator gives rise to 3, 4, 5, and 6-point weak-boson interactions. For the
process under consideration, it is the 3-point interactions that matter, namely
$\gamma W^+W^-$ and $Z W^+W^-$.  These interactions give rise to the Feynman
diagrams shown in Figure~\ref{fig:1}, which modify the differential cross section
for $f\bar f \to W^+W^-$. 

Our only information on this operator at present (at tree level) is from $f\bar
f \to W^+W^-$.  We can think of this operator as a modification of the triple
gauge boson interactions $\gamma W^+W^-$ and $Z W^+W^-$.  There is a standard
parameterization of anomalous triple gauge boson couplings that we can relate
this to \cite{Gaemers:1978hg,Hagiwara:1986vm}.  The relation is
\begin{equation}
\lambda_\gamma = \lambda_Z = c_{WWW}\frac{3g^2m_W^2}{2\Lambda^2}
\;,
\label{eq:lambda}\end{equation}
where $c_{WWW}/\Lambda^2$ is the coefficient of the operator ${\cal O}_{WWW}$ in
the Lagrangian. The fact that $\lambda_\gamma = \lambda_Z$ is a nontrivial
result that follows from restricting our attention to dimension-six operators.
If we were to include dimension-eight operators, the equality of these two
anomalous couplings would be violated \cite{Hagiwara:1993ck}.  Dimension-eight
operators are suppressed by $1/\Lambda^4$, so we expect these couplings to be
equal to good accuracy.

\begin{figure}[tb]
  \begin{center}
    \includegraphics[width=.9\linewidth]{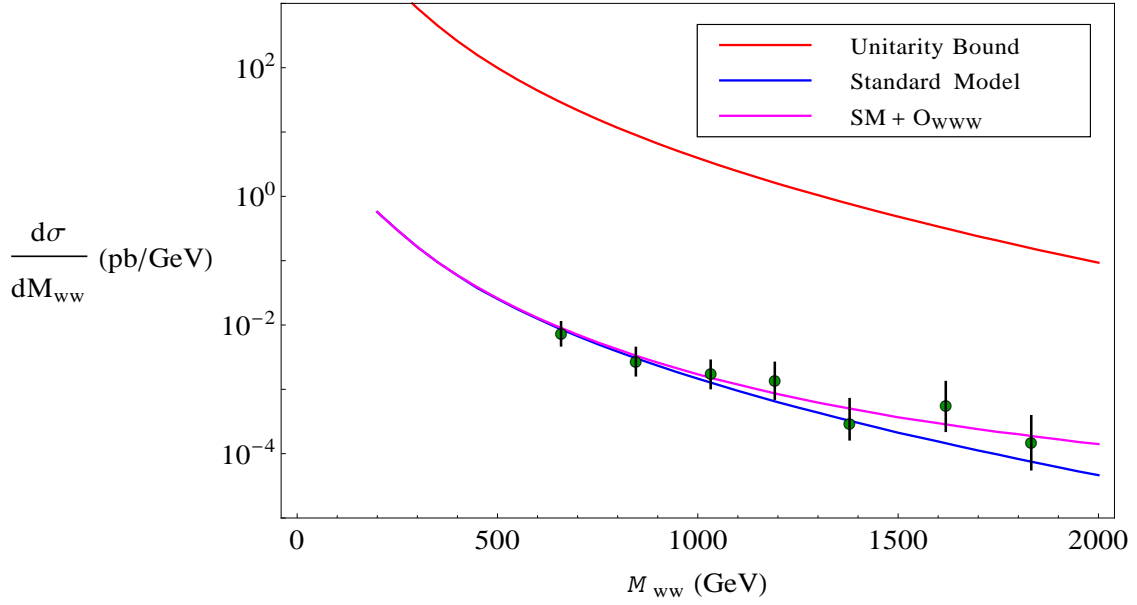}
  \end{center}
  \caption{The invariant mass distribution for $W^+W^-$ production at the LHC.
  The lowest (blue) curve is the Standard Model prediction; the middle (purple)
  curve includes the effect of the dimension-six operator ${\cal O}_{WWW}$;
  and the upper (red) curve is the unitarity bound.  The data are hypothetical.}
  \label{fig:2}
\end{figure}

It is well known that the presence of anomalous triple gauge boson couplings
leads to a cross section for $f\bar f \to W^+W^-$ that violates the unitarity
bound at high energy.  This led to the introduction of {ad hoc}
energy-dependent form factors such that the cross section respects the unitarity
bound at arbitrarily high energy \cite{Zeppenfeld:1987ip}.  The effective field
theory viewpoint makes it clear that this is an unnecessary complication
\cite{Degrande:2012wf}.  An effective field theory is only valid up to the scale
of new physics, $\Lambda$, not to arbitrarily high energy.  The data must
respect the unitarity bound, and since the effective field theory is only
intended to describe the data, it will also respect the unitarity bound.  It is
not a concern that the theoretical cross section violates the unitarity bound
beyond the energy where there is data.  This is exemplified in Figure~\ref{fig:2}.

The reason the dimension-six operator $O_{WWW}$, via $\lambda_{\gamma,Z}$
[Eq.~(\ref{eq:lambda})], leads to a cross section that violates the unitarity
bound at high energy, $E$, is that it introduces terms in the $f\bar f \to W^+W^-$
amplitude proportional to $c_{WWW}E^2/\Lambda^2$.
One method to place an upper
bound on $\Lambda$ is to calculate the energy at which the unitarity bound is
violated, for a given value of $c_{WWW}/\Lambda^2$.  This places an upper bound
on the energy at which the effective field theory is valid, and hence an upper
bound on $\Lambda$.  Such a calculation in the $V-A$ theory led to an upper
bound on the $W$ boson mass of about 700 GeV \cite{Abers:1973qs}.  The fact that
the $W$ boson mass is considerably less than this upper bound is a reflection of
the fact that the weak interaction is weakly coupled.

Unitarity bounds on anomalous triple gauge boson couplings should be viewed from
the same perspective \cite{Baur:1987mt,Gounaris:1994cm}.  For a given value of
an anomalous coupling, one can use the unitarity bound to calculate an upper
bound on the scale of new physics, $\Lambda$.  As the anomalous coupling tends
to zero, the upper bound on the scale of new physics recedes to infinity.

In the HISZ basis, there are several other operators that contribute directly to
the anomalous gauge boson couplings. Restricting our attention to the $CP$-even
operators, in addition to ${\cal O}_{WWW}$ there are
\begin{eqnarray}
{\cal O}_W & = & (D_\mu\Phi)^\dagger \hat{W}^{\mu\nu}(D_\nu\Phi) \label{eq:W}\\
{\cal O}_B & = & (D_\mu\Phi)^\dagger \hat{B}^{\mu\nu}(D_\nu\Phi) \label{eq:B}\\
{\cal O}_{BW} & = & \Phi^\dagger \hat{B}_{\mu\nu}\hat{W}^{\mu\nu}\Phi \label{eq:BW}\\
{\cal O}_{DW} & = & \tr\left( [D_\mu,\hat{W}_{\nu\rho}][D^\mu,\hat{W}^{\nu\rho}] \right)
\;.
\label{eq:DW}
\end{eqnarray}
The last operator contributes to $\lambda_\gamma = \lambda_Z$,\footnote{The
  operator ${\cal O}_{DW}$ also makes other contributions to the triple gauge
vertex.} while the first three operators contribute to the anomalous couplings
$\Delta g_1^Z$, $\Delta\kappa_Z$, $\Delta\kappa_\lambda$. However, one finds
that there is a relation amongst these anomalous couplings,
\begin{equation}
\Delta g_1^Z = \Delta\kappa_Z + s^2/c^2 \Delta\kappa_\gamma
\;,
\label{eq:g1z}\end{equation}
where $s,c$ are shorthand for $\sin\theta_W,\cos\theta_W$.  This relation is
again a consequence of restricting our attention to dimension-six operators; it
is violated by dimension-eight operators \cite{Hagiwara:1993ck}.

The relations of Eqs.~(\ref{eq:lambda},\ref{eq:g1z}) are examples of the way in
which an effective field theory gives some guidance as to what new interactions
to look for.  Another example is the absence of anomalous couplings amongst
neutral gauge bosons ($\gamma,Z$) when restricting ones attention to
dimension-six operators \cite{Hagiwara:1993ck,Degrande:2013kka}.

The two relations amongst the anomalous triple gauge couplings show that the
five couplings ($\lambda_Z, \lambda_\gamma, \Delta g_1^Z, \Delta\kappa_Z,
\Delta\kappa_\lambda$) are described by only three independent parameters.
However, the BW basis contains only two operators that contribute directly to
the triple gauge vertex, ${\cal O}_{W}$ and ${\cal O}_{\varphi WB}$
\cite{Han:2004az}.  This seems to imply that the anomalous triple gauge
couplings are described by only two independent parameters in this basis, which
apparently contradicts the claim that the physics is independent of the basis.
In particular, it appears that $\Delta g_1^Z=0$ in the BW basis.

The solution to this puzzle is that there are also indirect contributions to the
anomalous triple gauge couplings.  For example, in the BW basis the operator
(not listed in Table~\ref{tab:bases} since it includes fermions)
\begin{equation}
{\cal O}_{\varphi l}^{(3)}=\left(\varphi^\dagger i \overleftrightarrow{D}_\mu^I
\varphi\right)\left({\bar l}\tau^I\gamma^\mu l\right)
\end{equation}
affects the input parameter $G_F$ taken from muon decay, and this causes a shift
in $\sin^2\theta_W$, which in turn contributes to $\Delta g_1^Z$ and
$\Delta\kappa_Z$.  The relations of Eqs.~(\ref{eq:lambda},\ref{eq:g1z}) are
still respected.

In the GGPR basis, there are six operators that contribute directly to the
anomalous triple gauge couplings: ${\cal O}_{3W}$, ${\cal O}_{HW}$, ${\cal
O}_{HB}$, ${\cal O}_{2W}$, ${\cal O}_{B}$ and ${\cal O}_{W}$, which are the
same as the HISZ basis but with the HISZ operator ${\cal O}_{BW}$ replaced by
the GGPR operator ${\cal O}_B$ and ${\cal O}_{W}$.\footnote{
The operator ${\cal O}_{B}$ contributes by modifying the mixing of the $W^3$
and $B$ fields.}  One obtains the two relations amongst the anomalous triple
gauge couplings, Eqs.~(\ref{eq:lambda},\ref{eq:g1z}), demonstrating once again
that the same physics is obtained in any basis
\cite{Elias-Miro:2013mua,Pomarol:2013zra}.

\section{Flavor}
\label{sec:flavor}

Many dimension-six four-fermion operators contribute to processes in the
Standard Model that are forbidden at tree level and suppressed at one loop via
the GIM mechanism \cite{Glashow:1970gm}.  These processes put constraints on the
coefficients of these dimension-six operators that are of order $c_i/\Lambda^2
\sim (10^6$ GeV)$^{-2}$.  If the scale of new physics is $10^6$ GeV, there is
little hope of observing even its indirect effects at the LHC.  However, it is
possible that the coefficients, $c_i$, are much less than unity, in which case
$\Lambda$ could be much less than $10^6$ GeV.  Is there a natural explanation of
why flavor-violating dimension-six operators might have small coefficients while
flavor-conserving dimension-six operators have coefficients of order unity?

One such explanation goes under the name minimal flavor violation
\cite{Chivukula:1987py,Buras:2000dm,Buras:2003jf}, which can be incorporated
into the effective field theory approach
\cite{D'Ambrosio:2002ex,Grinstein:2007ef}.  The basic idea is that the
coefficients of the dimension-six operators that mediate flavor-changing
processes are suppressed by the same small factors that suppress these processes
in the Standard Model.

We will henceforth invoke minimal flavor violation as a rationale for
concentrating on flavor-conserving processes.  Minimal flavor violation suggests
that the largest flavor-violating effects are to be found in top quark physics.

\section{Beyond $S$ and $T$}
\label{sec:beyond}

A well-known parameterization of physics beyond the Standard Model is applicable
to models with heavy particles ($\gg m_Z$) that couple only to the electroweak
gauge bosons.  These particles contribute to the gauge boson self energies.  The
leading terms in an expansion of these self energies are called $S$, $T$, and
$U$ \cite{Peskin:1990zt}. 
A global fit to these parameters is performed by the
Gfitter group, and a plot of $S$ {vs.} $T$ (assuming $U=0$) may be
found in Ref.~\cite{Baak:2012kk} (see also Ref.~\cite{Gfitter}).

The effective field theory approach to physics beyond the Standard Model may be
viewed as a generalization of $S$ and $T$ to heavy particles that couple to more
than just the electroweak gauge bosons.  Using the conventions of
Ref.~\cite{Barbieri:2004qk}, the parameters $\hat S$ and $\hat T$ are
represented by the coefficients of the HISZ operators
\begin{eqnarray}
{\cal O}_{BW} & = & \Phi^\dagger \hat{B}_{\mu\nu}\hat{W}^{\mu\nu}\Phi
\\
{\cal O}_{\Phi,1} & = & (D_\mu\Phi)^\dagger\Phi \Phi^\dagger(D^\mu\Phi)
\;,
\end{eqnarray}
as
\begin{eqnarray}
\hat S & = & -\frac{c}{s}\Pi^\prime_{30}(0) = - c_{BW} \frac{m_W^2}{\Lambda^2}
\\
\hat T & = & - \frac{\Pi_{33}(0)-\Pi_{11}(0)}{m_W^2} = - c_{\Phi,1} \frac{v^2}{2\Lambda^2}
\;.
\end{eqnarray}
The parameter $\hat U$ does not correspond to any dimension-six operator.  It
first arises via a dimension-eight operator.

\begin{figure}[tb]
  \begin{center}
    \includegraphics[width=.7\linewidth]{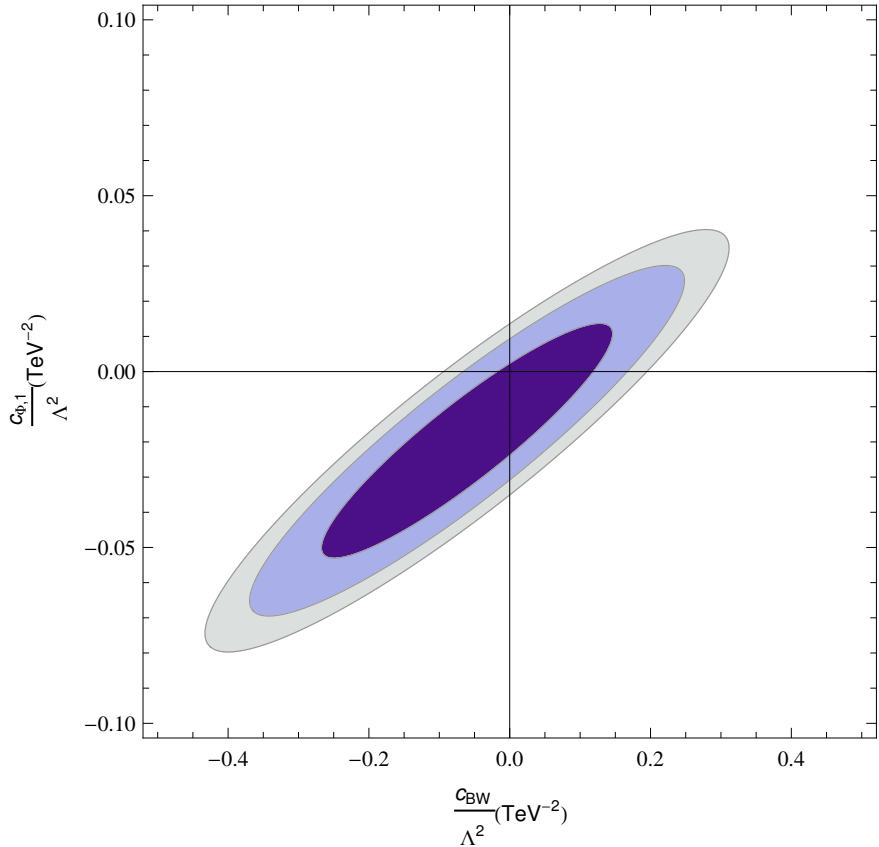}
  \end{center}
  \caption{Bound on $c_{BW}/\Lambda^2$ and $c_{\Phi,1}/\Lambda^2$, 
  assuming all other operator coefficients vanish.  The curves
  (from outer to inner) are 99\%, 95\% and 68\% confidence level.}
  \label{fig:3}
\end{figure}

Bounds on $\hat S$ and $\hat T$ yield bounds on $c_{BW}/\Lambda^2$ and
$c_{\Phi,1}/\Lambda^2$, as shown in Figure~\ref{fig:3}.  However, once one opens up the
parameter space to include all dimension-six operators, bounds on the coefficients
of these two operators become considerably looser
\cite{Han:2004az,Pomarol:2013zra,Grinstein:2013vsa}.

To elucidate this point, consider the plot of $c_{BW}/\Lambda^2$ {vs}.\
$c_{\Phi,1}/\Lambda^2$ in Figure~\ref{fig:3}. The strongest bound on
$c_{BW}/\Lambda^2$ is obtained by setting $c_{\Phi,1}/\Lambda^2 = 0$; however,
this is an {ad hoc} assumption.  Letting $c_{\Phi,1}/\Lambda^2$ float
loosens the bound on $c_{BW}/\Lambda^2$.  The loosest bound on
$c_{BW}/\Lambda^2$ is obtained by ignoring $c_{\Phi,1}/\Lambda^2$.  However,
this can be misleading, as it neglects the correlation between the bounds on
$c_{BW}/\Lambda^2$ and $c_{\Phi,1}/\Lambda^2$, captured by the ellipse in
Figure~\ref{fig:3}.

The analogous affects appear when we add more operators.  Letting the
coefficients of these operators float loosens the bounds on both
$c_{BW}/\Lambda^2$ and $c_{\Phi,1}/\Lambda^2$.  However, there continues to be a
correlation between $c_{BW}/\Lambda^2$ and $c_{\Phi,1}/\Lambda^2$, as well as a
correlation between these coefficients and the coefficients of the
additional operators.

In a different basis, $S$ and $T$ may be described by different operators.  In
the BW basis, they are described by the same operators as in the HISZ basis.  In
the GGPR basis, $T$ is described by a similar operator as in the other two
bases, but $S$ is given by the sum of the coefficients of two operators that
appear in neither the HISZ nor the BW basis,
\begin{eqnarray}
  {\cal O}_{W} & = & \frac{ig}{2}\left( H^\dagger\sigma^a \overleftrightarrow{D}^\mu
  H\right) D^\nu W^a_{\mu\nu} \\
  {\cal O}_{B} & = & \frac{ig'}{2}\left( H^\dagger \overleftrightarrow{D}^\mu H\right)
  \partial^\nu B_{\mu\nu}\;,
\end{eqnarray}
with\footnote{The operators in Ref.~\cite{Pomarol:2013zra} are suppressed by
inverse powers of $m_W$ instead of $\Lambda$, so the relation there is ${\hat
S}=c_W+c_B$.}
\begin{equation}
  {\hat S} = \frac{m_W^2}{\Lambda^2}\left(c_W + c_B\right)\;.
\end{equation}

It may be misleading to exclude consideration of a dimension-six operator under
the pretense that the coefficient of this operator is well constrained by
precision electroweak data.  A linear combination of such operators may be
equivalent, via equations of motion or other identities, to an operator that is
not well constrained \cite{Grojean:2006nn}.  If one includes all the operators
of a given basis in an analysis, this problem is avoided.  Operators should be
excluded only if it has been established that doing so will not bias an
analysis.

\section{Precision Electroweak Measurements}
\label{sec:pem}

Prior to the discovery of the Higgs boson, the most comprehensive analysis of
precision electroweak measurements using effective field theory
\cite{Grinstein:1991cd} was carried out in Ref.~\cite{Han:2004az}.  It included
every dimension-six operator that affects precision electroweak experiments, and
imposed flavor symmetry to exclude flavor-changing operators; it also imposed
$CP$.  This resulted in a list of 21 operators in the BW basis.  A global
analysis of the world's data was performed, resulting in a 
$\chi^2$ distribution as a function of 21 coefficients. The bounds on these
coefficients can be thought of as a 21-dimensional ellipsoid,
the analogue of the two-dimensional ellipse in the $c_{BW}/\Lambda^2$ {vs}.\
$c_{\Phi,1}/\Lambda^2$ plane of Figure~\ref{fig:3}.

As we discussed in Section~\ref{sec:beyond}, the bound obtained on the
coefficient of an operator by arbitrarily setting all other operator
coefficients to zero is too stringent. Conversely, the bound obtained on a
coefficient by letting all other coefficients float can be misleading, as the
correlation between coefficients is neglected. Both of these effects are
exacerbated by the large dimensionality of the space of coefficients.

Two linear combinations of BW operators would not be bounded at all if one were
to exclude the data from $e^+e^- \to W^+W^-$ \cite{Grojean:2006nn}.  These are
``blind directions'' \cite{De Rujula:1991se} with respect to the rest of the
precision electroweak data.  The two linear combinations of BW operators are
equivalent, via equations of motion, to the HISZ operators ${\cal O}_W$ and
${\cal O}_B$, or the GGPR operators ${\cal O}_{HW}$ and ${\cal O}_{HB}$.  In
this sense the BW basis is not as transparent as the HISZ or GGPR bases in
describing the precision electroweak data.  These two directions, as well as the
HISZ operator ${\cal O}_{WWW}$, are only weakly bounded by the $e^+e^- \to
W^+W^-$ data, which has limited statistics.

We have found that the $2^{\rm nd}$, $3^{\rm rd}$, and $7^{\rm th}$ smallest
eigenvalues of the bilinear function $\chi^2$ \cite{Han:2004az} have
eigenvectors which span these three weakly-bounded directions, to a good
approximation.  The eigenvectors with the $1^{\rm st}$, $4^{\rm th}$, and
$5^{\rm th}$ smallest eigenvalues span a linear combination of
quark-quark-lepton-lepton operators, and the eigenvector with the $6^{\rm th}$
smallest eigenvalue corresponds to a combination of four-lepton
operators.\footnote{We thank W.~Skiba for his assistance in this analysis.}

Because the directions corresponding to the HISZ operators ${\cal O}_W$ and
${\cal O}_B$ are present as linear combinations of operators in the BW basis,
it increases the correlations between operator coeffcients.  This is reflected
by the fact that the bound on the $S$ parameter is loosened by a factor of
about 2500 if all other operator coefficients are allowed to float {vs.}
setting them to zero \cite{Grinstein:2013vsa}.  In the GGPR basis, where these
directions are included in the basis via the operators ${\cal O}_{HW}$ and
${\cal O}_{HB}$, this factor is reduced to about 10 \cite{Pomarol:2013zra}.  A
similar result is expected in the HISZ basis.  While any basis can be used to
describe the data, the HISZ and GGPR bases have the advantage that the
weakly-bounded directions are included explicitly, which eliminates their
correlation with the precision electroweak data.

With the discovery of the Higgs boson, more dimension-six operators can now be
included. The analysis of Ref.~\cite{Han:2004az} sets the standard for future
analyses.

\section{Constraints at One Loop}
\label{sec:oneloop}

Some dimension-six operators are bounded only mildly.  For example, the HISZ
operators (also present in the BW basis; the second is also present in the GGPR
basis)
\begin{eqnarray}
{\cal O}_{WW} = \phi^\dagger W_{\mu\nu} W^{\mu\nu} \phi \\
{\cal O}_{BB} = \phi^\dagger B_{\mu\nu} B^{\mu\nu} \phi
\end{eqnarray}
are bounded at tree level only by measurements of the Higgs boson coupling to
electroweak bosons.  Prior to the discovery of the Higgs boson, these operators
were not bounded at all at tree level.  However, these couplings also contribute
to precision electroweak measurements at one loop.  It is conceivable that the
constraints on these operators from precision electroweak data are complementary
to those from tree-level Higgs phenomenology \cite{De
Rujula:1991se,Hagiwara:1992eh,Hagiwara:1993ck,Alam:1997nk,Elias-Miro:2013eta}.

\begin{figure}[tb]
  \begin{center}
    \includegraphics{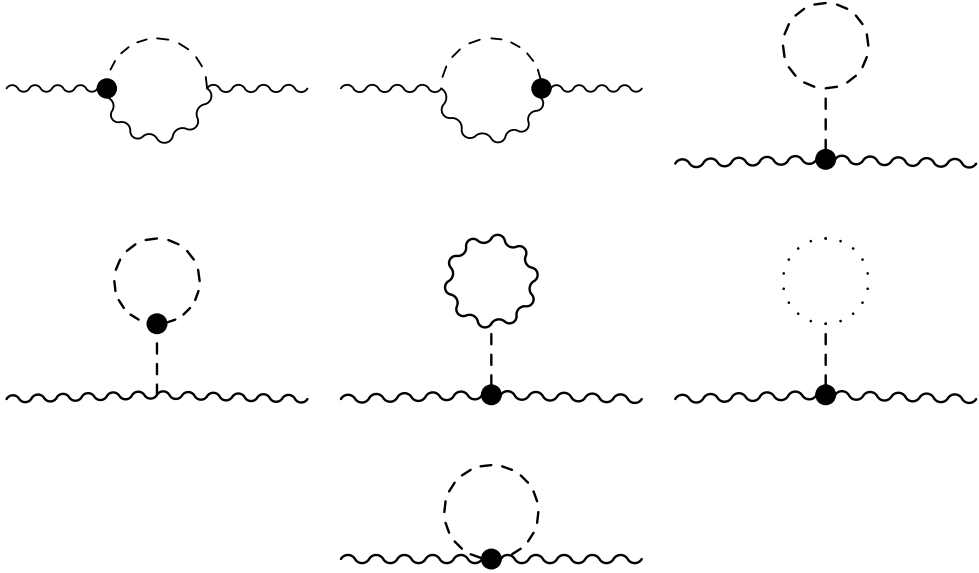}
  \end{center}
  \caption{Feynman diagrams involving operators ${\cal O}_{WW}$ and ${\cal
  O}_{BB}$ that contribute to precision electroweak measurements.
  The dots represent interactions arising from these operators ${\cal
  O}_{WW}$ and ${\cal O}_{BB}$.}
  \label{fig:4}
\end{figure}

The diagrams involving these operators that contribute to precision electroweak
measurements are shown in Figure~\ref{fig:4}.  Since the diagrams are all self energies
containing heavy particles, we can use the $S$ and $T$ formalism to describe the
result.  An explicit calculation reveals that the contribution to $\hat T$ vanishes,
while the contribution to $\hat S$ is ultraviolet divergent.  This means that the
HISZ operator ${\cal O}_{BW}$ must be included in the analysis.  Writing the
final expression in the large $m_h$ limit for simplicity, one finds
\cite{Mebane:2013cra,Mebane:2013zga,Chen:2013kfa}
\begin{equation}
\hat S = - \frac{c_{BW}(\mu)}{\Lambda^2}m_W^2 + \frac{g^2m_W^2}{32\pi^2}\left(\frac{c_{WW}}{\Lambda^2}+\frac{s^2}{c^2}\frac{c_{BB}}{\Lambda^2}\right)
\left[1-2\ln\left(\frac{m_h^2}{\mu^2}\right)\right]\label{eq:S}
\;,
\end{equation}
where $c_{BW}(\mu)$ is the renormalized coefficient of ${\cal O}_{BW}$ at the
renormalization scale $\mu$ in the $\overline{\rm MS}$ scheme.  Since this
coefficient is unknown, it screens the contribution of the operators ${\cal
O}_{WW}$ and ${\cal O}_{BB}$ to the $\hat S$ parameter.  In contrast, the $\hat
U$ parameter is ultraviolet finite, as it must be since it does not correspond
to any dimension-six operator:
\begin{equation}
\hat U = - \frac{g^2s^2m_W^2m_Z^2}{8\pi^2m_h^2}\frac{c_{WW}}{\Lambda^2}\;.
\end{equation}

Performing a global fit of the precision electroweak data to $c_{BW}$, $c_{WW}$,
and $c_{BB}$, yields only very weak constraints on the latter two coefficients
\cite{Mebane:2013cra,Mebane:2013zga,Chen:2013kfa}.  These constraints are much
weaker than originally thought \cite{Hagiwara:1993ck,Alam:1997nk}. Those
stronger bounds resulted from arbitrarily setting the renormalized coefficients
$c_{BW}(\mu)$ to zero, combined with setting the renormalization scale to $\mu =
\Lambda = 1$ TeV, which artificially enhances the one-loop contribution in
Eq.~(\ref{eq:S}) \cite{Mebane:2013zga}. An analysis of the one-loop contribution
of the HISZ operators ${\cal O}_{WWW}$, ${\cal O}_{W}$, and ${\cal O}_B$ to
precision electroweak data yields similarly weak constraints
\cite{Mebane:2013cra,Mebane:2013zga,Chen:2013kfa}. 

\section{Higgs boson}

The discovery of the Higgs boson has expanded the possibilities for probing
dimension-six operators.  This has led to a flurry of activity in applying
effective field theory to Higgs phenomenology
\cite{Manohar:2006gz,Passarino:2012cb,Dumont:2013wma,Chang:2013cia,Corbett:2012dm,
Corbett:2013pja,Einhorn:2013tja,Masso:2012eq,Contino:2013kra,
Elias-Miro:2013mua,Pomarol:2013zra,Alonso:2013hga}.  We anticipate that this
subject will continue to evolve, so we can only give a snapshot of the situation
as of Fall 2013.  The upcoming run of the LHC at 13 - 14 TeV will likely
fertilize further activity in this area.

The LHC data on Higgs production and decay are currently consistent with the
predictions of the Standard Model, so we can only put bounds on the coefficients
of dimension-six operators involving the Higgs boson. Ideally, this should be
done in the context of a global analysis, to include constraints on these
operators coming from other measurements. Starting with the 21 operators in the
BW basis needed to describe precision electroweak data identified in
Ref.~\cite{Han:2004az}, one needs to add several more operators to also describe
Higgs boson production and decay.  We count that an additional 7 operators are
needed, to bring the total to 28.  These are the BW operators ${\cal O}_{\varphi
W}$, ${\cal O}_{\varphi B}$, and ${\cal O}_{\varphi\Box}$ from
Table~\ref{tab:bases}; the operator
\begin{equation}
{\cal O}_{\varphi G} = \varphi^\dagger\varphi G_{\mu\nu}^A G^{A\mu\nu}
\end{equation}
to parameterize the Higgs coupling to gluons; and three operators to
parameterize the coupling of the Higgs boson to third generation fermions,
\begin{eqnarray}
{\cal O}_{u\phi} & = & \varphi^\dagger\varphi (\bar q u \tilde\varphi)\\
{\cal O}_{d\phi} & = & \varphi^\dagger\varphi (\bar q d \varphi)\\
{\cal O}_{e\phi} & = & \varphi^\dagger\varphi (\bar l e \varphi)\;.
\end{eqnarray}

At this time, there is no analysis that includes all 28 of these operators.  The
recent analyses
\cite{Passarino:2012cb,Dumont:2013wma,Chang:2013cia,
Corbett:2013pja,Einhorn:2013tja,Masso:2012eq,Contino:2013kra,
Elias-Miro:2013mua,Pomarol:2013zra} 
have been mostly concerned with describing the Higgs data, and include
other data to varying degrees.  One of the most extensive analyses at this time
includes 19 operators in the GGPR basis \cite{Pomarol:2013zra}. Of the GGPR
operators listed in Table~\ref{tab:bases}, the operators ${\cal O}_{2W}$, ${\cal
O}_{2B}$ are eliminated in favor of other operators using equations of
motion.\footnote{Ref.~\cite{Pomarol:2013zra} also lists the GGPR operator ${\cal
O}_6 = \lambda |H|^6$ in its basis, but this operator cannot be probed at
present.}

An analysis of Higgs and $W^+W^-$ data was performed in
Ref.~\cite{Corbett:2012ja} using six operators in the HISZ basis.  It was
pointed out in Ref.~\cite{Elias-Miro:2013mua} that this basis is incomplete,
because two of the operators that were removed via equations of motion
correspond to blind directions with respect to the precision electroweak data.
These operators were reinstated in a revised version of
Ref.~\cite{Corbett:2012ja} that is available on the arXiv.

Let's consider the decay $h \to \gamma\gamma$, which has the most statistical
significance. Because this decay proceeds at one loop in the Standard Model,
dimension-six operators that contributes to this decay are strongly constrained.
In the HISZ basis, the correction to the partial width for the decay receives
direct contributions from dimension-six operators proportional to
\begin{equation}
\Delta \Gamma(h\to \gamma\gamma) \sim (c_{WW} + c_{BB} - c_{BW})\;.
\end{equation}
The same formula applies in the BW basis, with the operator coefficients
appropriately renamed. In the GGPR basis,
\begin{equation}
\Delta \Gamma(h\to \gamma\gamma) \sim c_{BB}\;.
\end{equation}
In this basis it is evident that the bounds from $h\to \gamma\gamma$ are
independent of bounds from electroweak vector boson pair production, which
depend on the GGPR operators ${\cal O}_{3W}$, ${\cal O}_{HW}$, ${\cal O}_{HB}$,
${\cal O}_{2W}$, ${\cal O}_{B}$ and ${\cal O}_{W}$ \cite{Pomarol:2013zra}.

After $h\to \gamma\gamma$, the decays with the most significance are $h \to
Vf\bar f$, where $V=W,Z$ \cite{Grinstein:2013vsa,Isidori:2013cla}.  The
dimension-six operators that contribute to these decays are not independent of
those that contribute to electroweak vector boson pair production.  At present
the bounds on the operator coefficients from $h \to Vf\bar f$ and electroweak
vector boson pair production are comparable in magnitude
\cite{Pomarol:2013zra,Corbett:2013pja}.\footnote{Ref.~\cite{Corbett:2013pja}
uses an incomplete basis, as discussed above for Ref.~\cite{Corbett:2012ja}.
This is remedied in a revised version that is available on the arXiv.}  This is
less transparent in the BW basis, for reasons discussed in Sec.~\ref{sec:pem}.
The decay $h\to Z\gamma$, although it has not yet been observed, also bounds a
linear combination of the same operator coefficients.  In the GGPR basis
\cite{Pomarol:2013zra}
\begin{equation}
\Delta \Gamma(h\to Z\gamma) \sim \frac{1}{4}(c_{HW}-c_{HB})+2s^2c_{BB}
\end{equation}
which depends on the coefficients of the operators ${\cal O}_{HW}$, ${\cal
O}_{HB}$ that affect electroweak vector boson pair production.

The situation becomes more interesting if deviations from the Standard Model are
observed in the future.  By measuring the coefficients of the dimension-six
operators with good accuracy, we can hope to infer some or all of features of
the underlying theory.  The coefficients of the dimension-six operators are
measured at the electroweak scale, but the underlying theory resides at the
scale $\Lambda$, so it is best to evolve the measured coefficients from the
electroweak scale to the scale $\Lambda$ in order to get the most accurate
picture of the effective theory produced by the underlying physics.
Fortunately, theorists are hard at work deriving the necessary machinery for
this evolution
\cite{Grojean:2013kd,Elias-Miro:2013gya,Elias-Miro:2013mua,Jenkins:2013zja,Jenkins:2013wua,Jenkins:2013sda,Alonso:2013hga,Elias-Miro:2013eta}.
Unfortunately, as we mentioned previously, measuring the coefficients of the
dimension-six operators does not reveal the scale $\Lambda$, only the ratio
$c_i/\Lambda^2$.  Thus the coefficients must be evolved up to an arbitrary scale
$\Lambda$, unless we are able to deduce the scale $\Lambda$ by anticipating some
or all of the details of the underlying theory.  This is what happened with the
electroweak theory, as we recounted at the end of Section~\ref{sec:intro}.  It
remains to be seen whether this evolution is important to unraveling the
underlying physics.

To gain an understanding of the evolution of the operator coefficients, let's go
back to Eq.~(\ref{eq:S}).  Taking the derivative of this equation with respect
to the renormalization scale $\mu$ gives
\begin{equation}
\mu \frac{d}{d\mu}c_{BW}(\mu)= \frac{g^2}{8\pi^2}\left(c_{WW}+\frac{s^2}{c^2}c_{BB}\right)
\;,
\end{equation}
which is the evolution equation for the coefficient $c_{BW}(\mu)$; in general
there are also contributions to the right hand side from other operator
coefficients as well as Standard Model couplings.  The coefficients of the
operator coefficients on the right hand side of this equation are referred to as
the anomalous dimensions of $c_{BW}$, and the fact that the evolution depends on
operator coefficients different from $c_{BW}$ is referred to as operator mixing.
Due to operator mixing, the anomalous dimensions of the 59 operators is best
described by a matrix
\cite{Elias-Miro:2013mua,Jenkins:2013zja,Jenkins:2013wua,Elias-Miro:2013eta}.
The anomalous dimension matrix has recently been completed in the BW
basis \cite{Alonso:2013hga}.

\section{Final Thoughts}

We hope the reader is convinced that the effective field theory approach is a
simple and elegant way to treat new interactions beyond the Standard Model, with
many virtues.  It would be very exciting to discover new interactions and to use
effective field theory to help unravel the underlying physics.

We would like to end with a dose of humility.  Until a new collider is built, the
energy frontier belongs to the LHC.  Historically hadron colliders, as well as
proton fixed-target experiments, have discovered new physics by directly
observing new particles, not by observing the effective interactions that these
particles mediate at energies below their masses.  Notable examples include the
$J/\psi$, the $\Upsilon$, the $W$ and $Z$ bosons, the top quark, and the Higgs
particle.  While the $W$ and $Z$ bosons did manifest themselves first as
four-fermion interactions, these interactions were  observed in weak decays ($W$
boson) and in neutrino neutral currents ($Z$ boson), not in hadronic collisions.
The observation of new physics via its low-energy effective interactions is
unprecedented in hadronic collisions.  This can be viewed in a positive light;
after all, we would rather discover the new particles directly than observe
their low-energy effective interactions.

There are three facts behind this history.  First, all of these particles are
narrow resonances.  Second, hadronic collisions probe a broad range of
subprocess energies.  Third, there are generally large backgrounds in hadronic
collisions.  A narrow resonance sticks up above the background, if there is
enough subprocess energy to reach the resonance and enough parton luminosity to
produce the resonance. The effective physics at energies below the resonance is
not observable.

There is another historical point we would like to make.  We have already
recounted the story that led from the Fermi theory to the $V-A$ theory to the
electroweak model, and cited this as evidence of the usefulness of effective
field theory.  However, consider another effective field theory, that of the
low-energy interactions of pions and kaons known as chiral perturbation theory
\cite{Weinberg:1978kz}.  Although we now know that the underlying theory is QCD,
chiral perturbation theory continues to be a useful tool.  However, it was not
chiral perturbation theory that led to the discovery of QCD.  That discovery
came via an entirely different route, including the observation of scaling in
deep inelastic scattering and the theoretical realization of asymptotic freedom
in non-Abelian gauge theories.  Thus the discovery and measurement of effective
interactions, while important, may not be the thing that allows us to deduce the
underlying theory.

\section*{Acknowledgements}

We are grateful for conversations and correspondence with A.~Manohar, C.~Quigg,
and W.~Skiba. S.~W.\ is supported in part by U.S. Department of Energy grant
DE-FG02-13ER42001. C.~Z.\ is supported by the IISN ``Fundamental interactions''
convention 4.4517.08.

\bibliographystyle{unsrtnat}

\end{document}